# OVERVIEW OF THE NSTX CONTROL SYSTEM*

P. Sichta, J. Dong, G. Oliaro, P. Roney, PPPL, Princeton, NJ 08543, USA

Abstract

The National Spherical Torus Experiment (NSTX) is an innovative magnetic fusion device that was constructed by the Princeton Plasma Physics Laboratory (PPPL) in collaboration with the Oak Ridge National Laboratory, Columbia University, and the University of Washington at Seattle. Since achieving first plasma in 1999, the device has been used for fusion research through an international collaboration of over twenty institutions. The NSTX is operated through a collection of control systems that encompass a wide range of technology, from hardwired relay controls to real-time control systems with giga-FLOPS of capability. This paper presents a broad introduction to the control systems used on NSTX, with an emphasis on the computing controls, data acquisition, and synchronization systems.

## 1 INTRODUCTION

NSTX achieved first plasma in February, 1999. At the time, only ten computing-control subsystems were in use. Today, there are over fifty control subsystems. A diagram of the NSTX control system architecture is shown in fig. 1. The control subsystems can be divided into two categories, 'engineering' and 'physics'. Engineering control systems are used to create and control the fusion reactions that occur in the plasma. Physics systems, also known as diagnostics, are used to gather information about the plasma.

The NSTX device is operated in a 'pulsed' mode. Typically, the duration of an NSTX fusion pulse (a shot) is about 0.5 seconds. The torus' extremely compact center stack assembly needs about ten minutes to cool down between pulses.

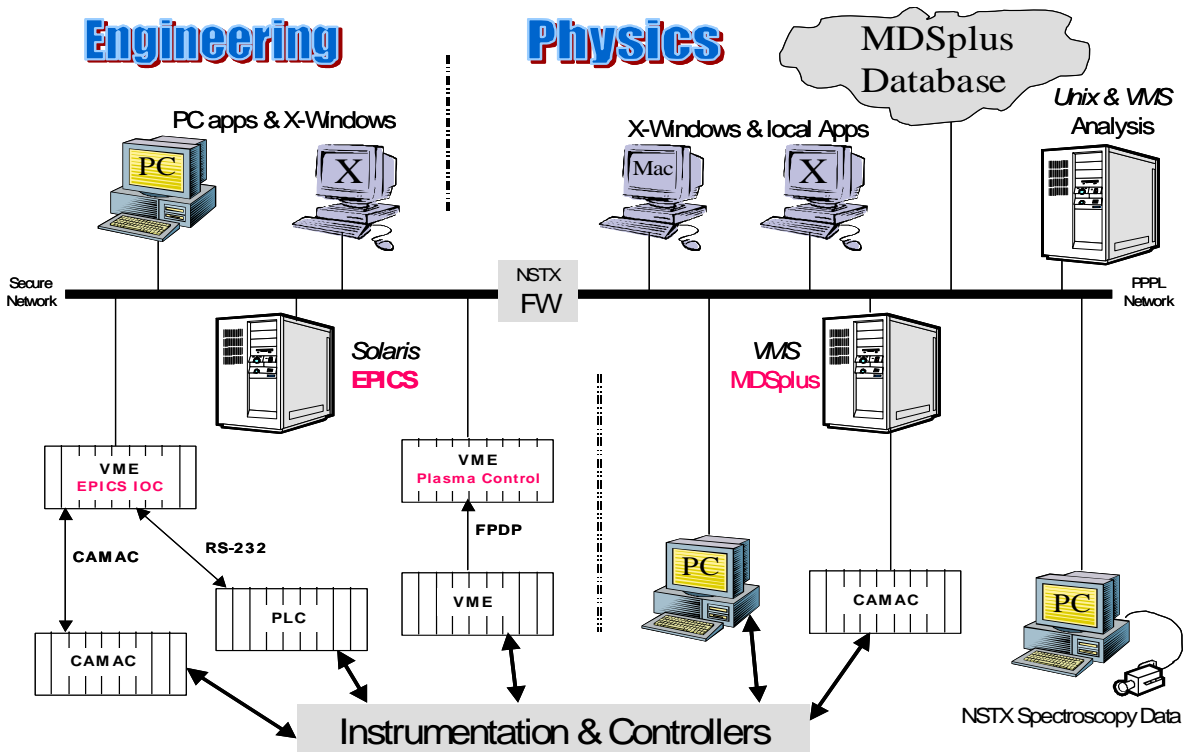

Fig 1 – Architecture of the NSTX Control System. There are Engineering subsystems and Physics subsystems. A variety of operating systems and human interface devices are supported.

Typically, the control systems record data at high-speed into 'local' memory during the short plasma pulse. The data is subsequently deposited into the *MDSplus* data management system. Physics analysis codes produce data plots and other scientific results, that can be used to 'tweak' the control systems for the next shot.

The NSTX control system is made up of commercial and collaboratively-developed components. Examples of the hardware include VME single board computers, x-terminals, workstations, and personal computers. Operating systems include Solaris, VMS, MacOS, Windows, and vxWorks. The bulk of the application software is *collaborative software*. This software is open-source, in use on a variety of projects, and is enhanced and maintained through the community of its users.

## 2 ENGINEERING SYSTEMS

The engineering systems provide the functionality needed to operate NSTX. Examples of these subsystems are the torus vacuum system, the water-cooling system, and the field coil power supplies. The systems incorporate a variety of hardware technologies. Functions that involve personnel and equipment protection typically employ older, well-characterized technologies such as relay logic or hardwired analog processing systems. Non-protective controls often use advanced technology, such as VME, to provide high performance and flexible operation. Programmable logic controllers (PLC) provide a middle ground, offering high reliability and a modern human-machine interface. Most NSTX engineering subsystems are interfaced with the Experimental Physics and Industrial Control System (EPICS), which has served NSTX well in its first two years of operation [1].

The most advanced control system at NSTX is the plasma control system[3]. This system controls the shape, position, and density of the plasma. It is a real-time digital control system that acquires over 200 input signals. Each millisecond, the field coil power supplies and gas injection systems are sent new control commands. The system features a 160 MB/s Front Panel Data Port (FPDP), an 8-processor, 20 giga-FLOP SkyBolt II computer, and real-time fiber optic data links. The system uses the Plasma Control System software, developed by General Atomics, Inc. for the DIII-D tokamak.

## 3 PHYSICS SYSTEMS

The physics control systems, or diagnostics, measure properties of the plasma. Most diagnostics are passive, in that they cannot affect the plasma. A few provide real-time signals representing a critical plasma parameter, such as plasma density. This signal can be used as a feedback signal for other control systems.

An increasing number of diagnostics are designed, operated, and maintained by non-PPPL scientists at collaborating institutions. PPPL liaison engineers help interface the system with NSTX facility. Considerations include mechanical interfaces, power and grounding, timing, and data management. The latter is very important since high level physics analysis codes sometimes require data from several diagnostics. New diagnostics often use a standard desktop PC. In addition to being inexpensive, the equipment that is connected to the PC often has software provided by the equipment's manufacturer. These attributes can significantly lower the overall cost of the diagnostic.

## 4 NSTX SOFTWARE

The NSTX software was designed to use commercial and collaborative software, i.e. to minimize new system software development. Collaborative software has permitted NSTX computer support to be highly effective and reliable. Collaborative software packages used on NSTX are the Experimental Physics and Industrial Control System (EPICS), the Model Data System (MDSplus)[2], Scope, the Plasma Control System software (PCS), and the Inter-Process Communication System (IPCS).

*4.1 EPICS*
The EPICS software is used to integrate most engineering subsystems. It is extensible to many I/O technologies and runs under several popular operating systems. EPICS is used to acquire and archive NSTX engineering data. The NSTX-specific application is presently comprised of 3 IOCs, 135 EPICS process displays, 1,500 I/O points, and 7,000 EPICS records. EPICS is designed for continuous process control applications. 'C' programs and unix-scripts were required to support the special needs of NSTX's 'pulsed' operations and to interface EPICS with other NSTX software.

The EPICS human machine interface is implemented using over twenty x-terminals and PCs. These are used by the operators to monitor all engineering subsystems and, if authorized, control them.

*4.2 MDSplus*
NSTX data management is based on MDSplus, which is a hierarchically organized database that is created for each NSTX shot. Both raw and analyzed data from the engineering and physics control systems are written into the 'shot tree'. Currently, MDSplus stores 100 MB of data per shot, more than 5000 waveforms and 25,000 parameters. The NSTX MDSplus server runs on VMS. MDSplus clients run under VMS, unix, and Windows. There are FORTRAN, C, IDL, and Java interfaces to the MDSplus database.

In addition to data storage, MDSplus provides device control capability and an event system. For example, an MDSplus event can be used to refresh a plot with new data, or to control a CAMAC-interfaced instrument.

*4.3 Visualization*

The main data visualization tools on NSTX are *Scope* and *IDL*. Scope is a 2-D data visualization tool that can plot from one to sixty four signals in an X-window. The tool is designed to access data from the MDSplus shot tree. It has a convenient graphical user interface with the capability to pan and zoom any or all X-Y plots. In response to MDSplus events, it can automatically re-plot a new shot's data from the MDSplus shot tree. Scope runs on unix and VMS.

IDL is often used for complex data analysis and visualization. IDL functions are used to interface with MDSplus. IDL is also used as a basis for an NSTX Physics electronic logbook.

*4.4 Web Tools*

Many NSTX users interact with the NSTX control systems and data using application programs that use the X-windows interface. New tools are increasingly aimed towards the world-wide-web browser interface. NSTX uses a web interface for several purposes. First, there are 'help' functions such as FAQs, links to local and collaborators' documentation, and training presentations. There are search tools to help the user find signals in the MDSplus database, and plotting and data-export tools. The logbook also has a web-browser interface. Some of the new MDSplus tools use Java.

## 5 TIMING AND SYNCHRONIZATION

All control systems use the NSTX Central Clock system for timing and synchronization. The system encodes up to fifteen pre-defined timing events, with one microsecond resolution. The Central Clock hardware is made up of CAMAC modules. There is real-time software running under EPICS, and several EPICS displays to configure and control the clock.

Software synchronization at NSTX is based on two software packages, the Inter-Process Communication System (IPCS) and the Event Manager (EVT). IPCS runs on unix and VMS and has both C and FORTRAN application program interfaces. EVT is an event distribution system that is built on top of IPCS. These messaging programs are used to coordinate software (tasks) that are executing on different computers. EPICS is used to interface the Central Clock hardware with these programs.

## 6 CONCLUSION

The ability to quickly produce and maintain a highly reliable control system by a small group of developers would not be possible without generous support from our collaborators at institutions around the world. The use of collaborative software has helped to make a software environment that spans multiple operating systems and programming languages. Active collaborations also provide an avenue to keep the software compatible with modern computing technologies.

The NSTX control system is rapidly expanding to include new and diverse systems. The trend of new engineering and diagnostic subsystems is toward independent system design and development by 'the experts' in the applicable area of research or engineering, not by a central NSTX engineering group. A key requirement for these systems is their ability to interface with the MDSplus data management system, the Central Clock system, and in the case where integrated systems control is required, with EPICS.

To maintain the reliability of the control system that was built using legacy CAMAC equipment, a program to replace aging equipment with modern hardware will need to be completed. Exploratory work to improve the NSTX Central Clock to provide additional performance and capability is underway. Emerging technologies such as CompactPCI and FPGA devices are currently being explored.